\documentstyle[epsfig]{mn}

\title{CDM models with a steplike initial power spectrum}
\author[Mirt Gramann and Gert H\"utsi] 
{Mirt Gramann and Gert H\"utsi 
   \\ 
   Tartu Observatory,
       T\~oravere 61602, Estonia}
\begin{document}
\maketitle

\let\sec=\section 
\let\ssec=\subsection 
\let\sssec=\subsubsection

\def\kms{\;{\rm km\,s^{-1}}}
\def\kmsmpc{\;{\rm km\,s^{-1}\,Mpc^{-1}}}
\def\hompc{\,h\,{\rm Mpc}^{-1}}
\def\mpcoh{\,h^{-1}\,{\rm Mpc}}
\def\mpc3h{\,h^{3}\,{\rm Mpc^{-3}}}

\begin{abstract}
We investigate the properties of clusters of galaxies in the
$\Lambda$CDM models with a steplike initial power spectrum. We examine
the mass function, the peculiar velocities and the power spectrum of
clusters in models with different values of the density parameter $\Omega_0$,
the normalized Hubble constant $h$ and the spectral parameter $p$, which
describes the shape of the initial power spectrum. The results are
compared with observations. We also investigate 
the rms bulk velocity in the models, where the properties of clusters
are consistent with the observed data. We find that the power spectrum of 
clusters is in good agreement with the observed power spectrum of the Abell-ACO
clusters, if the spectral parameter $p$ is in the range $p=0.6-0.8$. The
power spectrum and the rms peculiar velocity of clusters are consistent
with observations only if $\Omega_0<0.4$. The $\Omega_0=0.3$ models are 
consistent with the observed properties of clusters, if $h=0.50-0.63$.
For $h=0.65$, we find that $\Omega_0=0.20 -0.27$. 

\end{abstract}

\begin{keywords}
cosmology: theory -- large-scale structure of Universe,
cosmology: theory -- dark matter, galaxies: clusters
\end{keywords}

\sec{INTRODUCTION}

The theory of galaxy formation based on gravitational instability 
describes how primordially generated fluctuations grow into galaxies 
and clusters of galaxies due to self-gravity of matter. The initial field 
of density fluctuations $\delta({\bf x},t)$ 
can be decomposed into its Fourier components $\delta_{\bf k}(t)$ and 
expressed in terms of the power spectrum 
$P(k)=\langle\vert\delta_{\bf k} \vert^2\rangle$.

As the study of the large-scale structure in the Universe has pushed to ever 
larger scales, several data samples have suggested the presence of a peak 
in the power spectrum at the wavenumber $k \simeq 0.05h$ Mpc$^{-1}$ (or at the 
wavelength $\lambda \simeq 120 - 130 h^{-1}$ Mpc). 
Einasto et al. (1997) and Retzlaff et al. (1998) studied the
spatial distribution of the Abell-ACO clusters and found that the power
spectrum of the clusters has a well-defined peak at the 
wavenumber $k=0.052h$ Mpc$^{-1}$. A similar peak in the one-dimensional 
power spectrum of a deep pencil-beam survey was detected by 
Broadhurst et al. (1990), and in the two-dimensional power spectrum of 
the Las Campanas redshift survey by Landy et al. (1996). 
The power spectrum of the spatial distribution of APM galaxies also has
a feature on the same scale (Caztanaga and Baugh 1998).
Independent
evidence for the presence of a preferred scale in the Universe at about
$130h^{-1}$ Mpc comes from an analysis of high-redshift galaxies.
Broadhurst and Jaffe (1999) studied the distribution of the Lyman-break
galaxies at redshift $z \sim 3$ and found a $5\sigma$ excess of pairs
separated by $\Delta z=0.22 \pm 0.02$, equivalent to $130h^{-1}$ Mpc for
flat universe with the density parameter $\Omega_0=0.4 \pm 0.1$.

The power spectrum of density fluctuations depends on
the physical processes in the early universe. The peak in the power
spectrum of clusters at the wavenumber $k \simeq 0.05h$ Mpc$^{-1}$ may
be generated during the era of radiation domination or earlier.
Standard cosmological models based on collisionless dark matter 
[e.g. cold dark matter (CDM)] and adiabatic fluctuations, when combined 
with power-law initial power spectra, predict smooth power spectra of 
density fluctuations at $z\sim 10^3$. The baryonic acoustic oscillations 
in adiabatic models may explain the observed power spectrum only if 
currently favored determinations of
cosmological parameters are in substantial error (e.g., if the density
parameter $\Omega_0<0.2h$; Eisenstein et al. 1998). 

\begin{figure*}
\centering
\begin{picture}(300,350)
\includegraphics{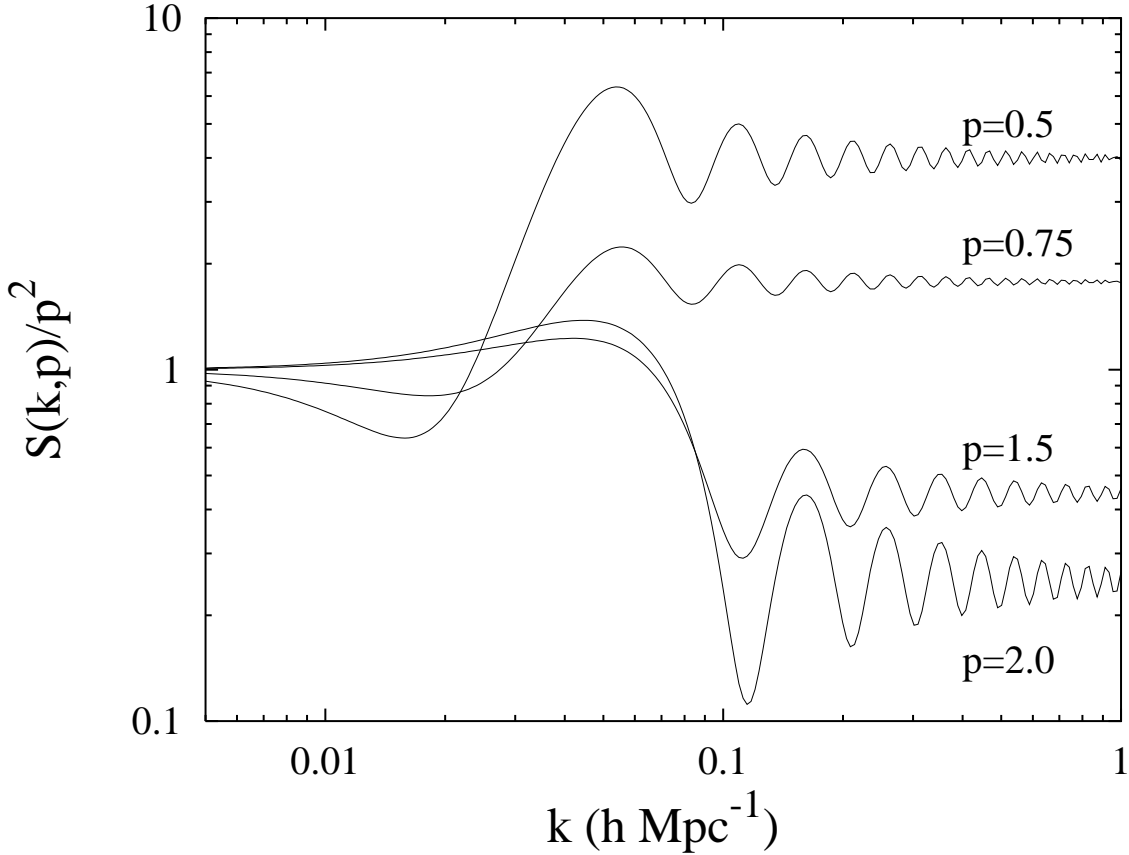}
\end{picture}
\caption {The function $S(k,p)/p^2$ for different values of the 
parameter $p$. For the models with $p<1$ and $p>1$, the step parameter 
was chosen to be $k_0=0.016h$ Mpc$^{-1}$ and $k_0=0.03h$ Mpc$^{-1}$, 
respectively.}
\end{figure*}

One possible explanation for the observed power spectrum of clusters is an 
inflationary model with a scalar field whose potential $V(\varphi)$ has a 
local steplike feature in the first derivative. This feature can be produced 
by a fast phase transition in a physical field different from the inflaton 
field. An exact analytical expression for the scalar (density) perturbations 
generated in this inflationary model was found by Starobinsky (1992). 
The initial power spectrum of density fluctuations in this model can be 
expressed as
$$
P_{in}(k) \propto {k \, S(k,p) \over p^2} ,
\eqno(1)
$$
where the function $S(k,p)$ can be written as
$$
S(k,p) = 1 -
{3 (p-1) \over y}\left[f_1(y)\sin{2y} + {2 \over y} \cos{2y}\right] +
$$
$$
+{9 \, (p-1)^2 \, f_2(y) \over 2 \, y^2}
       \left[f_2(y)+f_1(y)\cos{2y} - {2 \over y} \sin{2y} \right] \, .
\eqno(2)
$$
Here, the function $y=k/k_0$, $f_1(y)=1-y^{-2}$ and $f_2(y)=1+y^{-2}$.
The initial power spectrum in this model depends on two 
parameters $k_0$ and $p$. The parameter $k_0$ determines the location of 
the step and the parameter $p$ - the shape of the initial spectrum.
For $p=1$, we recover the scale-invariant Harrison-Zel'dovich
spectrum ($S(k,1) \equiv 1$). At present, the initial spectrum (1,2)
is probably the only example of an initial power spectrum with the
desired properties, for which a closed analytical form exists.

Fig.~1 shows the function $S(k,p)/p^2$ for different values of
the parameter $p$. For the models with $p<1$ and $p>1$, the step parameter
was chosen to be $k_0=0.016h$ Mpc$^{-1}$ and $k_0=0.03h$ Mpc$^{-1}$, 
respectively. In this case, in the models with $p<1$, the power spectrum 
has a well-defined maximum at the wavenumber $k\simeq 0.05h$ Mpc$^{-1}$ 
and a second maximum at $k \simeq 0.1h$ Mpc$^{-1}$. In the models with 
$p>1$, the picture is inverted. The power spectrum has a flat upper 
plateau at the wavenumbers $k<0.05h$ Mpc$^{-1}$, a sharp decrease on smaller 
scales ($k=0.05-0.1h$ Mpc$^{-1}$) and a secondary maximum at
$k \simeq 0.15h$ Mpc$^{-1}$. 

Lesgourgues, Polarski \& Starobinsky (1998, hereafter LPS) compared the 
CDM models with a steplike initial spectrum (1,2) with 
observational data. They studied the rms mass fluctuation on an 
$8h^{-1}$ Mpc scale, $\sigma_8$; the rms bulk velocity of galaxies and the 
cosmic microwave background (CMB) anisotropies on different angular scales. 
In this paper we continue this research and investigate the properties of 
galaxy clusters in these models. We examine the mass function, the peculiar 
velocities and the power spectrum of clusters 
for different values of the density parameter $\Omega_0$,
the normalized Hubble constant $h$ and the spectral parameter $p$.
The results are compared with observations. We also 
investigate the rms bulk velocity and $\sigma_8$ in the 
models, where the mass function, the peculiar velocities and the power
spectrum of clusters are consistent with the observed data. 

In their study LPS assumed that the parameter
$\sigma_8$ lies in the interval 
$\sigma_8=(0.57 \pm 0.06) \Omega_0^{-0.56}$. This interval was derived by
White, Efstathiou \& Frenk (1993) by analysing the mass function of
clusters. For $\Omega_0=0.3$ this gives $\sigma_8=1.0-1.25$. In this
paper we examine the observed values of the mass function of galaxy
clusters in more detail and obtain lower values for the parameter
$\sigma_8$. For $\Omega_0=0.3$ we find that $\sigma_8=0.8-1.05$. 
Therefore, the allowed values for the parameter $p$ in the 
($\Omega_0$, $h$) plane that we obtain are different from those 
found by LPS.

We examine flat cosmological models with the density parameter 
$\Omega_0=0.2-0.5$ and the normalized Hubble constant $h=0.5-0.8$.
These parameters are in agreement with measurements of the density 
parameter (e.g. Bahcall et al. 1999) and with measurements of the Hubble 
constant using various various distance indicators (e.g. Tammann 1998). 
To restore the spatial flatness in the low-density models, we assume a 
contribution from a cosmological constant: $\Omega_{\Lambda}=1-\Omega_0$.
The Hubble constant is written as $H_0=100h$ km s$^{-1}$Mpc$^{-1}$. 
The transfer function $T(k)$ is computed with 
the fast Boltzmann code CMBFAST developed by Seljak \& Zaldarriaga (1996).
The models are normalized using the COBE normalization derived by Bunn
\& White (1997). We assume that the initial density fluctuation field
in the Universe is a Gaussian field. In this case, the power spectrum
provides a complete statistical description of the field.

This paper is organized as follows. In Section 2 we study the mass
function of clusters of galaxies and compare the
results with observations. In Section 3 we examine the peculiar velocities 
of galaxy clusters. In Section 4 we investigate the power 
spectrum of clusters. Discussion and summary are presented in Section 5.

\sec{THE MASS FUNCTION OF CLUSTERS OF GALAXIES}

To study the mass function of clusters we use the Press-Schechter 
(1974, PS) approximation. The PS mass function has been compared with
N-body simulations (Efstathiou et al. 1988; White, Efstathiou \& Frenk 
1993; Lacey \& Cole 1994; Eke, Cole \& Frenk 1996; Borgani et al. 1997a) 
and has been shown to provide an accurate description of the abundance 
of virialized cluster-size halos. In the PS approximation the number 
density of clusters with the mass between $M$ and $M+dM$ is given by
$$
n(M) dM = - \sqrt{{2 \over \pi}} {\rho_b \over M} 
{\delta_t \over \sigma^2(M)} {d\sigma(M)\over dM}  
\exp \left[-{\delta_t^2 \over 2\sigma^2(M)}\right]  dM .  
\eqno(3)
$$
Here $\rho_b$ is the mean background density and $\delta_t$ is the 
linear theory overdensity for a uniform spherical fluctuation which is
now collapsing; $\delta_t=1.686$ for $\Omega_0=1$, with a weak dependence
on $\Omega_0$ for flat and open models (Eke et al. 1996; Kitayama \&
Suto 1996). In this paper we will use values of $\delta_t$ found by Eke
et al. (1996) for flat models.
The function $\sigma(M)$ is the rms linear density fluctuation at the
mass scale $M$. We will use the top-hat window function to describe halos.
For the top-hat window, the mass $M$ is related to the window radius
$R$ as $M=4\pi\rho_b R^3/3$. In this case, the number density of clusters 
of mass larger than $M$ can be expressed as
$$
n_{cl}(>M) = \, {\int_M^{\infty}} n(M') dM' = 
$$
$$ 
= - {3 \over (2\pi)^{3/2}} {\int_R^{\infty}} { \delta_t \over \sigma^2(r)} 
{d\sigma(r)\over dr} \exp\left[-{\delta_t^2 \over 2\sigma^2(r)}\right] 
{dr \over r^3}  .   
\eqno(4)
$$

Fig.~2 shows the cluster mass function in the flat models with 
$\Omega_0=0.3$ and $h=0.65$. The threshold density $\delta_t=1.675$. We 
investigated the cluster masses within a $1.5h^{-1}$ Mpc radius sphere 
around the cluster center. This mass 
$M_{1.5}$, is related to the window radius $R$ as 
$$
R=8.43 \Omega_0^{0.2\alpha \over 3-\alpha} \left[{M_{1.5} \over 6.99
\times 10^{14} \Omega_0 h^{-1} M_{\odot}}\right]
^{1 \over 3-\alpha} (h^{-1} {\rm Mpc})  .
\eqno(5)
$$
Here the parameter $\alpha$ describes the cluster mass profile, 
$M(r) \sim r^\alpha$, at radii $r\sim 1.5h^{-1}$ Mpc. Numerical 
simulations and observations of clusters indicate that the parameter 
$\alpha \approx 0.6-0.7$ for most of clusters  (Navarro, Frenk \& 
White 1995; Carlberg, Yee \& Ellingson 1997). In this paper we use a 
value $\alpha=0.65$. 

\begin{figure*}
\centering
\begin{picture}(300,350)
\includegraphics{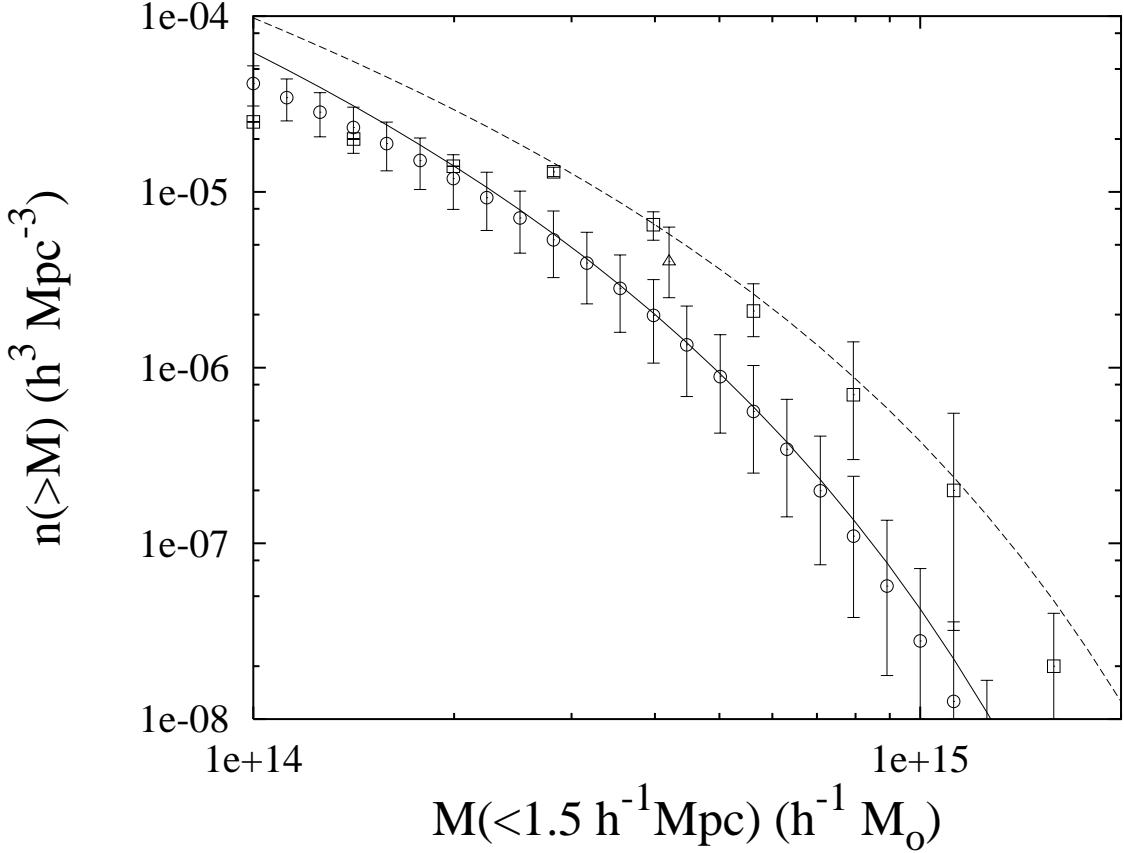}
\end{picture}
\caption {The cluster mass function in the models with 
$\Omega_0=0.3$, $h=0.65$, $p=1.0$ (solid line) and $p=0.79$ (dashed line).  
Open circles and squares show the mass function of galaxy clusters 
derived by Bahcall and Cen (1993) and by Girardi et al. (1998), 
respectively. The open triangle is the result obtained by White, 
Efstathiou \& Frenk (1993).}
\end{figure*} 

Fig.~2 shows also the mass function of clusters of galaxies derived
by Bahcall and Cen (1993, BC) and by Girardi et al. (1998, G98). BC 
used both optical and X-ray observed properties of clusters to determine 
the mass function of clusters. The function was extended towards the faint 
end using small groups of galaxies. G98 determined the mass function 
of clusters by using virial mass estimates for 152 nearby 
Abell-ACO clusters including the ENACS data (Katgert et al. 1998). The 
mass function derived by G98 is somewhat larger than the mass function 
derived by BC, the difference being larger at larger masses (see Fig.~2). 

Let us consider the amplitude of the mass function of galaxy clusters at 
$M_{1.5} = 4\cdot 10^{14} h^{-1} M_{\odot}$. For this mass, the cluster 
abundances derived by BC and G98 are $n(>M)=(2.0 \pm 1.1)\cdot 10^{-6} 
h^3$ Mpc$^{-3}$ and $n(>M)=(6.3\pm 1.2)\cdot 10^{-6} h^3$ Mpc$^{-3}$, 
respectively. By analysing X-ray properties of clusters, White, Efstathiou 
\& Frenk (1993) found that the number density of clusters with the mass 
$M_{1.5} \approx 4.2 \cdot 10^{14} h^{-1} M_{\odot}$ is 
$n(>M)=4 \cdot 10^{-6} h^3$ Mpc$^{-3}$. 

We derived the limits for the parameter $p$, assuming that the mass function 
of galaxy clusters at $M_{1.5} = 4\cdot 10^{14} h^{-1} M_{\odot}$ is in 
the range $(2-6.5) \cdot 10^{-6} h^3$ Mpc$^{-3}$. Fig.~3a shows the 
results for the models with $h=0.65$ for different $\Omega_0$ and Fig.~3b 
for the models with $\Omega_0=0.3$ for different $h$. For the model with 
$\Omega_0=0.3$ and $h=0.65$, we find that 
$p=0.79-1.0$. Fig.~2 demonstrates the mass function of clusters in 
this model for $p=0.79$ and $p=1.0$.

LPS used high values of the parameter $\sigma_8$ and found that 
one of the best-fit models is the model with parameters 
$\Omega_0=0.3$, $h=0.7$, $p=0.8$. We find that the number density of
clusters in this model is substantially higher than 
observed; for the mass $M_{1.5} = 4\cdot 10^{14} h^{-1} M_{\odot}$,
the cluster abundance $n(>M)=9.8 \times 10^{-6} h^3$ Mpc$^{-3}$. For 
$\Omega_0=0.3$ and $h=0.7$, the mass function of clusters is consistent 
with the observed data, if $p=0.88-1.13$.
\begin{figure*}
\centering
\begin{picture}(300,490)
\includegraphics{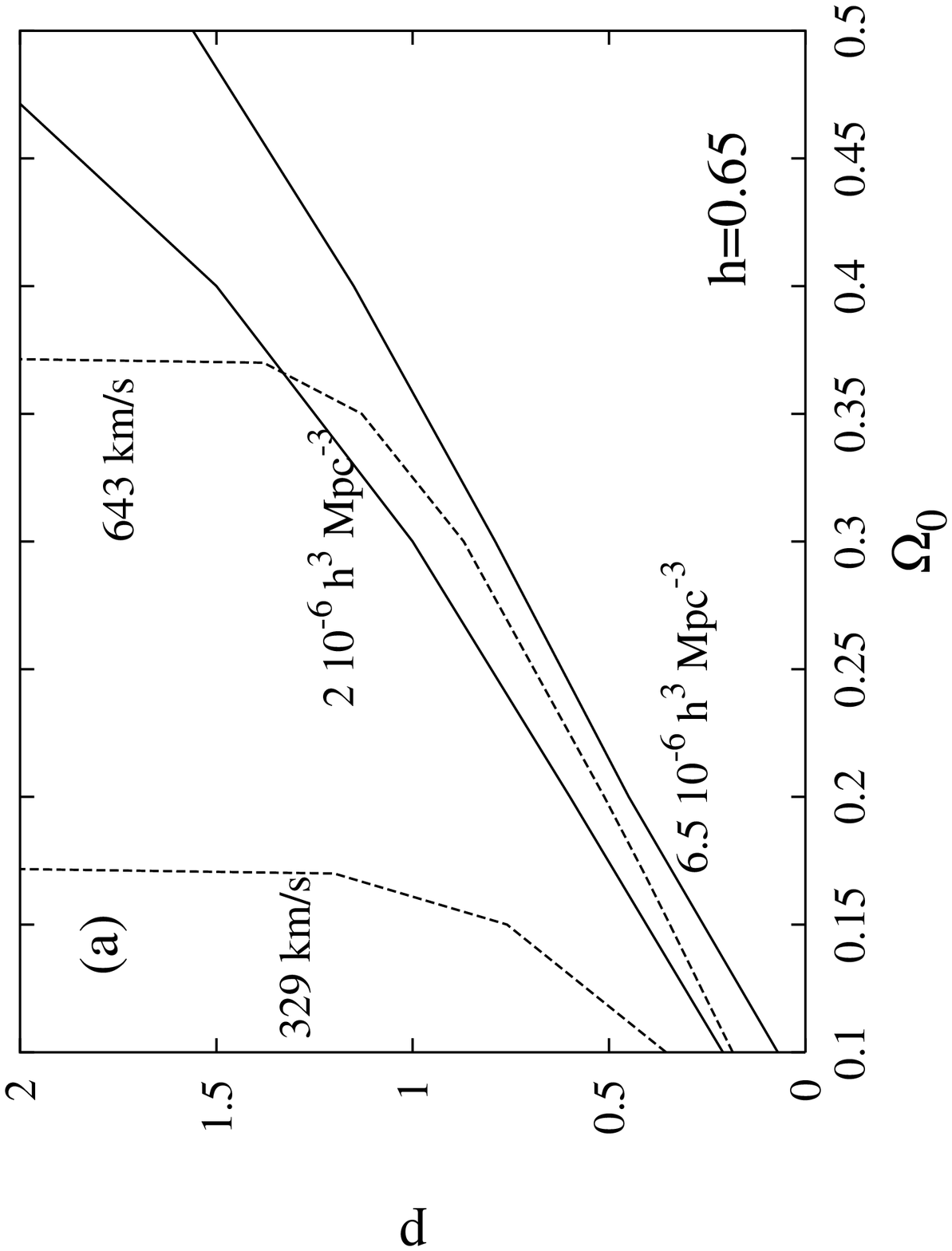}
\includegraphics{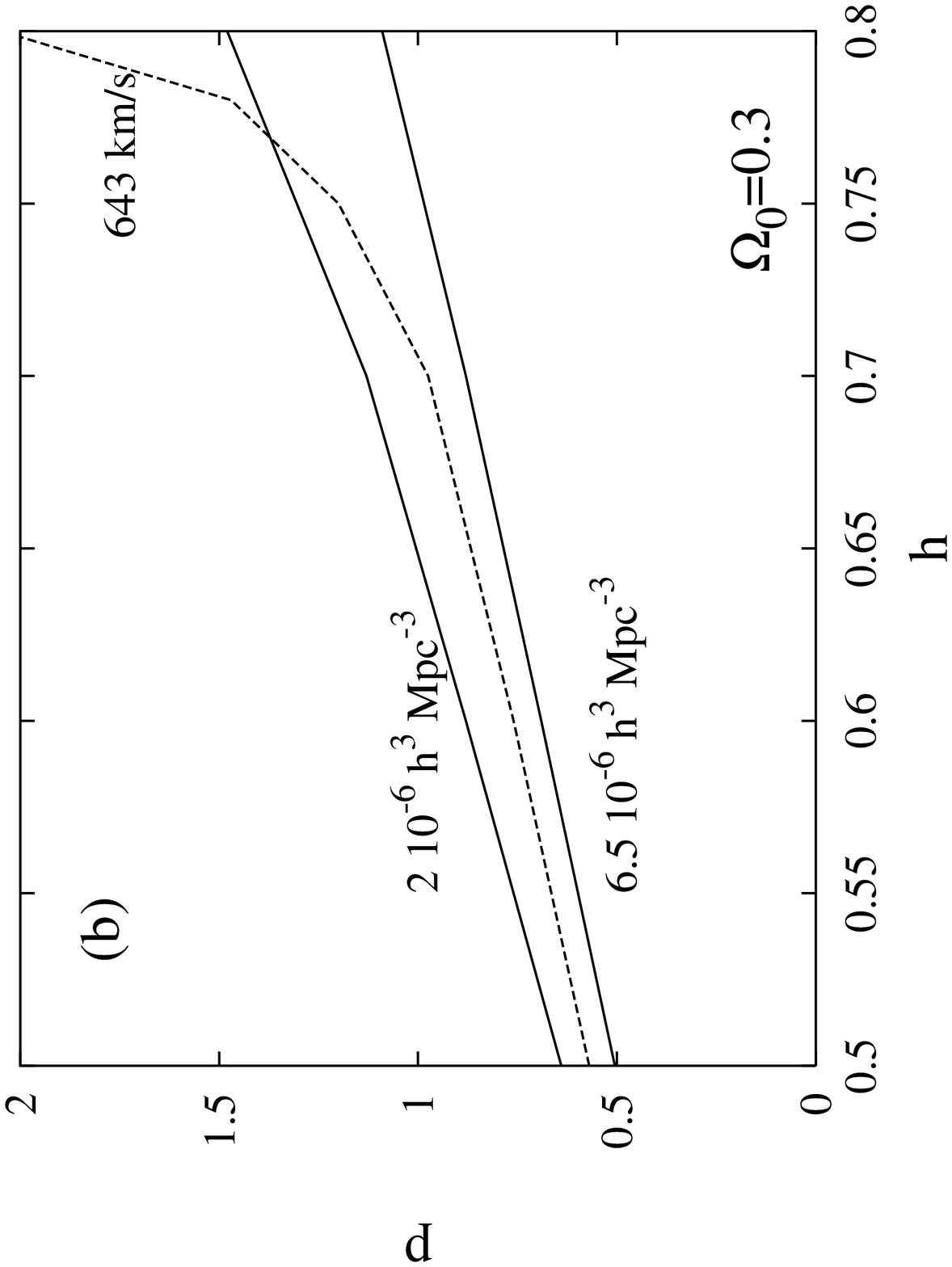}
\end{picture}
\caption {The limits for p in the $\Lambda$CDM models with a steplike 
initial power spectrum. Solid lines show the constraints obtained from 
the mass function of clusters and dashed lines show the 
constraints obtained by analysing the peculiar velocities of clusters. 
(a) $h=0.65$. (b) $\Omega_0=0.3$. In the region studied peculiar 
velocities are larger than $329$ km s$^{-1}$.}
\end{figure*} 

\sec{PECULIAR VELOCITIES OF CLUSTERS OF GALAXIES}

The observed rms peculiar velocity of galaxy clusters has been studied 
in several papers (e.g. Bahcall, Gramann \& Cen 1994, 
Bahcall and Oh 1996, Borgani et al. 1997b, Watkins 1997). In this paper
we use the results obtained by Watkins (1997). He developed a
likelihood method for estimating the rms peculiar velocity of clusters
from line-of-sight velocity measurements and their associated errors.
This method was applied to two observed samples of cluster peculiar 
velocities: a sample known as the SCI sample (Giovanelli et al. 1997) and 
a subsample of the Mark III catalog (Willick et al. 1997).  Watkins (1997) 
found that the rms one-dimensional cluster peculiar velocity is 
$265^{+106}_{-75}$ km s$^{-1}$, which
corresponds to the three-dimensional rms velocity  
$459^{+184}_{-130}$ km s$^{-1}$.

To investigate the peculiar velocities of clusters in our models,
we use the linear theory predictions for peculiar 
velocities of peaks in the Gaussian field. The linear rms velocity 
fluctuation on a given scale $R$ at the present epoch can be expressed 
as $$
\sigma_v(R)=H_0 f(\Omega_0) \sigma_{-1} (R),
\eqno(6)
$$
where $f(\Omega_0) \approx \Omega_0^{0.56}$ is the linear
velocity growth factor in the flat models and  $\sigma_j$ is defined for 
any integer $j$ by 
$$
\sigma_j^2={1 \over 2\pi^2} \int P(k) W^2(kR) k^{2j+2} dk.
\eqno(7)
$$
Bardeen et al. (1986) showed that 
the rms peculiar velocity at peaks of the smoothed density field differs 
systematically from $\sigma_v(R)$, and can be expressed as
$$
\sigma_p(R)=\sigma_v(R) \sqrt{1 - \sigma_0^4/\sigma_1^2 \sigma_{-1}^2} .
\eqno(8)
$$

Suhhonenko \& Gramann (1999) examined the linear theory predictions 
for the peculiar velocities of peaks and compared these to the peculiar
velocities of clusters in N-body simulations. The N-body clusters were 
determined as peaks of the density field smoothed on the scale 
$R \sim 1.5h^{-1}$ Mpc. The numerical results
showed that the rms peculiar velocity of small clusters is similar to
the linear theory expectations, while the rms peculiar velocity of rich
clusters is higher than that predicted in the linear theory.
The rms peculiar velocity of clusters with a mean cluster separation 
$d_{cl} = 30h^{-1}$ Mpc was $\sim 18$ per cent higher than that predicted 
by the linear theory. We assume that the observed cluster sample studied by 
Watkins (1997) corresponds to the model clusters with a separation 
$d_{cl}\sim 30h^{-1}$ Mpc ($n_{cl} \sim 3.7 \cdot 10^{-5} h^3$ Mpc$^{-3}$)
and determine the rms peculiar velocity of the 
clusters, $v_{cl}$, as
$$
v_{cl}=1.18 \, \sigma_p(R),
\eqno(9)
$$
where the radius $R=1.5h^{-1}$ Mpc.  

Fig.~3 shows the limits for $p$ in different models obtained on the basis
of Watkins (1997) results. Fig.~3a shows the results in the $h=0.65$ models 
for different $\Omega_0$ and Fig.~3b in the $\Omega_0=0.3$ models for
different $h$. In the region studied in Fig.~3b, the rms
peculiar velocity is larger than 329 km s$^{-1}$. For the model with 
$\Omega_0=0.3$ and $h=0.65$, we find that the initial parameter $p>0.87$. 

For high values of $\Omega_0$ and $h$, peculiar velocities are
larger than observed for any values of the parameter $p$. 
The velocities are sensitive to the amplitude of the large-scale
fluctuations at wavenumbers $k<0.1h$ Mpc$^{-1}$. By increasing $p$, if 
$p>1$, the power spectrum on these wavenumbers and, therefore, the 
velocities remain almost unchanged (see Fig.~1). Fig.~3a shows that
in the $h=0.65$ models with $\Omega_0>0.35$, the peculiar velocities are 
not consistent with observations. For the $\Omega_0=0.4$ model with $h=0.5$ 
and $h=0.6$, we find that peculiar velocities are consistent with 
observations, if $p>0.86$ and $p>1.26$, respectively.

If we compare the observational constraints obtained by studying
the mass function and peculiar velocities of clusters of galaxies, we
see from Fig.~3 that the mass function and the peculiar velocities of clusters 
are consistent with the observed data only in a small interval of the 
parameter $p$. For the model with $\Omega_0=0.3$ and $h=0.65$, we find 
that $p=0.87-1.0$. For the $\Omega_0=0.3$ models with $h=0.5$ and
$h=0.6$, we find that $p=0.57-0.64$ and $p=0.76-0.88$, respectively. 
 
\sec{THE POWER SPECTRUM OF CLUSTERS OF GALAXIES}

Let us now consider the power spectrum of clusters in these models where
both the mass function and the peculiar velocities are consistent with
observations. To investigate the power spectrum of clusters,
we can also use the PS approach. The power spectrum of clusters for a given 
number density in the PS approximation can be expressed as (Gramann \&
Suhhonenko 1999, hereafter GS)
$$
P_{cl}(k)=b_{cl}^2 \, P(k),
\eqno(9)
$$
where the cluster bias parameter $b_{cl}$ is
$$
b_{cl} = 1 - {3 \over (2\pi)^{3/2}n_{cl}} {\int_R^{\infty}}
{ 1 \over \sigma^2(r)} \, {d\sigma(r)\over dr}
(y^2 -1) \exp (-y^2) {dr \over r^3} .
\eqno(10)
$$
Here, the function $y=\delta_t/ \sigma(r)$ and $\sigma(r)$ is the rms
linear density fluctuation at the radius $r$. 
The cluster bias parameter $b_{cl}$ depends on the minimal mass $M$ (or
the window radius $R$) of clusters and on the power spectrum of density
fluctuations, $P(k)$, which determines the function $\sigma(r)$. For
fixed $P(k)$ and $n_{cl}$, the minimal mass $M$ (or scale $R$) can be
determined by inverting equation (4). 

Observations provide the distribution of clusters in redshift
space, which is distorted due to peculiar velocities of clusters. On
large scales, where linear theory applies, the power spectrum of matter
density fluctuations in redshift space is given by (Kaiser 1987):
$$
P^s(k)=\left[1 + {2 f(\Omega_0) \over 3} + {f^2(\Omega_0)
\over 5} \right] P(k).
\eqno(11)
$$
Using the PS approximation (10), relation (11)
takes the form
$$
P^s_{cl}(k)=
\left[1 + {2 f(\Omega_0) \over 3 b_{cl}} + {f^2(\Omega_0) \over
5 b_{cl}^2} \right] b_{cl}^2 \, P(k).
\eqno(12)
$$
Equation (12) determines the power spectrum of clusters
for a given $n_{cl}$ in redshift space.

GS examined the power spectrum of clusters in the PS theory and in N-body 
simulations. They determined the power spectrum of clusters for mean 
separations $d_{cl}=30-40h^{-1}$ Mpc. The numerical results showed that at
wavenumbers $k<0.1h^{-1}$ Mpc, the power spectrum of clusters in the
simulations is linearly enhanced with respect to the power spectrum of
the matter distribution. However, the amplitude of the spectrum of
clusters was somewhat lower than predicted by the approximation 
(12). It is possible that the linear approximation 
(12) overestimates the power spectrum of clusters due to dynamical
effects that are not taken into account in this approximation. 
The power spectrum of clusters can be expressed as
$$
P^s_{cl}(k)= 
\, F \,\left[1 + {2 f(\Omega_0) \over 3 b_{cl}} + {f^2(\Omega_0) \over
5 b_{cl}^2} \right] b_{cl}^2 \, P(k),
\eqno(13)
$$
where the factor $F=0.7-0.8$. The factor $F$ depends slightly on the
model. GS examined the distribution of clusters in two cosmological
models which start from the observed power spectra of the distribution
of galaxies and clusters of galaxies. In the model (1), the initial
linear power spectrum of density fluctuations was chosen in the form
$P(k) \propto k^{-2}$ at wavelengths $\lambda<120h^{-1}$ Mpc. In the
model (2), GS assumed that the initial power spectrum contains a
primordial feature at the wavelengths $\lambda \sim 30-60h^{-1}$ Mpc.
They found that $F=0.8$ and $F=0.7$ in the model (1) and model (2), 
respectively. In this paper we use a value $F=0.75$.

Fig.~4 shows the redshift-space power spectrum of the model clusters 
with a mean separation $d_{cl}=34h^{-1}$ Mpc. For comparison, we show
the power spectrum of the Abell-ACO clusters determined by Einasto et
al. (1999). This spectrum represents the weighted mean of the power
spectra determined by Einasto et al. (1997) and Retzlaff et al. (1998).
Einasto et al. (1997) determined the power spectrum of the Abell-ACO
clusters from the correlation function of clusters, while Retzlaff et
al. (1998) estimated the power spectrum directly (see Einasto et al.
1999 for details). The power spectrum of the distribution of the
Abell-ACO clusters peaks at the wavenumber $k=0.052h$ Mpc$^{-1}$. The
mean intercluster separation of the Abell-ACO clusters is $d_{cl} \sim
34h^{-1}$ Mpc ($n_{cl} \sim 2.5 \cdot 10^{-5} h^3$ Mpc$^{-3}$)
(Einasto et al. 1997, Retzlaff et al. 1998). 

\begin{figure*}
\centering
\begin{picture}(300,350)
\includegraphics{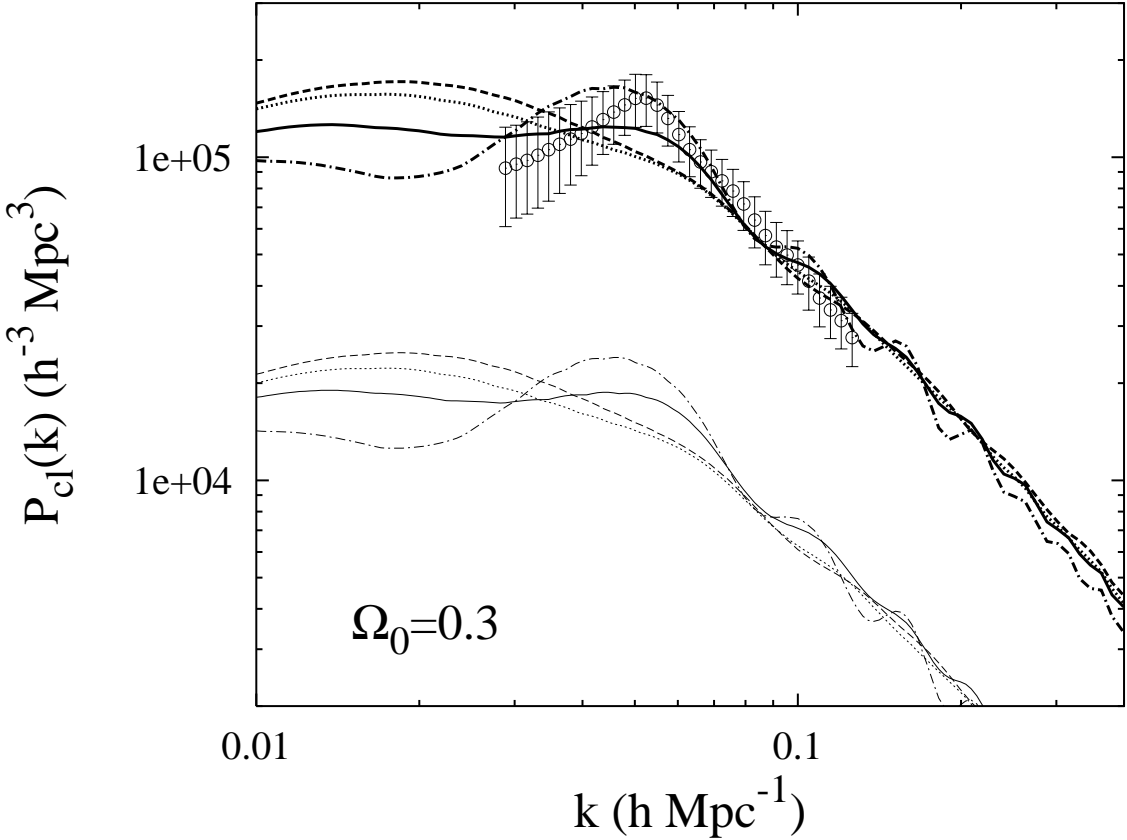}
\end{picture}
\caption {The power spectrum of the model clusters and the Abell-ACO
clusters (open circles) with a separation $d_{cl}=34h^{-1}$ Mpc. 
The heavy lines show the power spectra of clusters determined by
approximation (13) and the light lines the corresponding linear power
spectra of matter fluctuations. We have shown the spectra for the 
$\Omega_0=0.3$ models with $h=0.5$, $p=0.6$ (dot-dashed lines), 
$h=0.6$, $p=0.8$ (solid lines), $h=0.65$, $p=0.95$ (dotted lines) and 
$h=0.7$, $p=1.05$ (dashed lines).}
\end{figure*} 

Fig.~4 shows the power spectrum of clusters predicted in the
$\Omega_0=0.3$ models with ($h=0.5$, $p=0.6$), ($h=0.6$, $p=0.8$), 
($h=0.65$, $p=0.95$) and ($h=0.7$, $p=1.05$). For these models, the 
mass function and the peculiar velocities are consistent with the
observed data (see Fig.~3b). The models with $p=0.6$ and $p=0.8$  
are in good agreement with the observed power spectrum, while
the models with $p=0.95$ and $p=1.05$ are not consistent with the
observed data. In the models with $p>0.95$, we do not see a peak in the 
power spectrum at the wavenumber $k\simeq 0.05h$ Mpc$^{-1}$. 

We used also the $\chi^2$ test to calculate the probability that the
models fit the observed power spectrum. The observed power spectrum
represents the weighted mean of the power spectra determined by Einasto
et al. (1997) and Retzlaff et al. (1998) and it is not clear how many
points of the power spectrum are independent. As a first step we
used all the data points in Fig.~4. The models with $p=0.6$ and $p=0.8$ 
are consistent with the observed power spectrum at a confindence level 
higher than $90$\%. For the models with $p=0.95$ and $p=1.05$, the 
probability that the models fit the observed power spectrum is less than 
$30$\%. 

We studied also the power spectrum of clusters in other models 
with different values of ($\Omega_0$, $h$) and found similar results. The 
power spectrum of clusters is in good agreement with the observed power 
spectrum of the Abell-ACO clusters, if the initial parameter $p$ is in 
the range $p=0.6-0.8$. In the models with $h=0.65$, the allowed values 
for the parameter $p$ are in this range, if $\Omega_0=0.20-0.27$ 
(see Fig.~3a). Similarly, the $\Omega_0=0.3$ models are consistent 
with the observed mass function, peculiar velocities and the power spectrum
of clusters, if $h=0.50-0.63$ (Fig.~3b). 

In the $\Omega_0=0.4$ model with $h=0.5$, the peculiar velocities are larger 
than observed, if $p<0.86$. For higher values of $\Omega_0$
and $h$, this limit for $p$ increases. Therefore, for $\Omega_0 \geq 0.4$ and 
$h \geq 0.5$, the observed power spectrum and peculiar velocities of clusters 
are not consistent with each other. Either the observed peculiar velocities 
are underestimated, or the observed peak in the power spectrum of clusters 
is overestimated. The peculiar velocities and 
the power spectrum of clusters are consistent with observations only if 
$\Omega_0<0.4$.   

\sec{DISCUSSION AND SUMMARY}

In this paper, we have examined the properties of clusters of galaxies
in the $\Lambda$CDM models with a steplike initial power spectrum (1,2)
that depends on two parameters $k_0$ and $p$. The parameter $k_0$
determines the location of the step and the parameter $p$ - the shape
of the initial spectrum. For the models with $p<1$ and $p>1$, the step
parameter was chosen to be $k_0=0.016h$ Mpc$^{-1}$ and 
$k_0=0.03h$ Mpc$^{-1}$, respectively.  We investigated the mass function, 
peculiar velocities and the power spectrum of clusters in models with 
different values of $\Omega_0$, $h$ and $p$. 

We found that the mass function and the peculiar velocities of clusters 
are consistent with the observed data only in a small interval of the 
parameter $p$. For the model with $\Omega_0=0.3$ and $h=0.65$, we find 
that $p=0.87-1.0$. For the $\Omega_0=0.3$ models with $h=0.5$ and
$h=0.6$, we find that $p=0.57-0.64$ and $p=0.76-0.88$, respectively. 

The power spectrum of clusters in the $\Lambda$CDM models with a
steplike initial power spectrum is in good agreement with the observed 
power spectrum of the Abell-ACO clusters, if the initial parameter $p$ is 
in the range $p=0.6-0.8$. In the models with $h=0.65$, the allowed values 
for the parameter $p$ are in this range if $\Omega_0=0.20-0.27$. 
The $\Omega_0=0.3$ models are consistent with the observed mass function, 
peculiar velocities and the power spectrum of clusters, if $h=0.50-0.63$.
The peculiar velocities and the power spectrum of clusters are consistent 
with observations only if $\Omega_0<0.4$.   

We also studied the rms bulk velocity in the
models, where the mass function, the peculiar velocities and the power 
spectrum of clusters are consistent with observations. We choosed three 
models in the allowed parameter space and studied these models in more detail. 
Tabel~1 lists the parameters of the models studied. We have determined the
number density of clusters with mass 
$M_{1.5} = 4\cdot 10^{14} h^{-1} M_{\odot}$, $n_{cl}$;
the rms mass fluctuation on an $8h^{-1}$ Mpc scale, $\sigma_8$; 
the rms peculiar velocity of clusters, $v_{cl}$, and the rms bulk
velocity for a radius $r=50h^{-1}$ Mpc, $V_{50}$. The rms bulk velocity
was determined by using eq. (6). The observed
bulk velocities are determined in a sphere centered on the Local Group
and represent a single measurement of the bulk flow on large scales.
The observed bulk velocity derived from the Mark III catalog of peculiar
velocities for $r=50h^{-1}$ Mpc is $375 \pm 135$ km s$^{-1}$ (Kolatt \&
Dekel 1997). In the models studied, the rms bulk velocity is 
$\sim 270-285$ km s$^{-1}$, which is consistent
with the observed data. 

Therefore, in many aspects the $\Lambda$CDM models with a steplike
initial spectrum fit the observed data. Further work is needed to study 
the properties of galaxies and clusters of galaxies in these models in 
more detail.

\sec*{ACKNOWLEDGEMENTS}

We thank A. Starobinsky, J. Einasto and E. Saar for useful discussions. 
This work has been supported by the ESF grant 3601.

\begin{table*}
\begin{minipage}{120mm}
\caption{Results of the tests for the different models.}
\begin{tabular}{|c|c|c|c|c|c|c|}
$\Omega_0$ & $h$ & $p$ & $n_{cl}$ & $\sigma_8$ & $v_{cl}$ & $V_{50}$ \\
    &   &   & ($10^{-6}h^3$ Mpc$^{-3}$) &  & (km s$^{-1}$) & (km s$^{-1}$) \\
\hline
0.20 & 0.75 & 0.70 & 2.9 & 1.05 & 610 & 270 \\ 
0.25 & 0.65 & 0.75 & 2.8 & 0.95 & 610 & 270 \\
0.30 & 0.60 & 0.80 & 3.3 & 0.90 & 635 & 285 \\

\end{tabular}
\label{table}
\end{minipage}
\end{table*}

\vfill


\begin{thebibliography}{}

\bibitem[\protect\citename{Bahcall}{1993}]{Bahcall93}
Bahcall, N.A., Cen, R. 1993, ApJ, 407, L49 (BC)
\bibitem[\protect\citename{Bahcall}{1994}]{Bachall94}
Bahcall, N.A., Gramann, M., Cen, R. 1994, ApJ, 436, 23 
\bibitem[\protect\citename{Bahcall}{1996}]{Bahcall96}
Bahcall, N.A., Oh, S.P. 1996, ApJ, 462, L49
\bibitem[\protect\citename{Bahcall}{1999}]{Bahcall99}
Bahcall, N.A., Ostriker, J.P., Perlmutter, S., Steinhardt, P.J. 1999,
Science, 1481 
\bibitem[\protect\citename{Bardeen}{1986}]{Bardeen96}
Bardeen, J.M., Bond, J.R., Kaiser, N., Szalay, A.S. 1986,
ApJ, 304, 15
\bibitem[\protect\citename{Borgani}{1997a}]{Borgani97a}
Borgani, S. et al. 1997a, NewA, 1, 321
\bibitem[\protect\citename{Borgani}{1997b}]{Borgani97b}
Borgani, S., Da Costa L.N., Freudling, W., Giovanelli, R., Haynes,
M.P., Salzer, J., Wegner, G., 1997b, ApJ, 482, L121 
\bibitem[\protect\citename{Broadhurst}{1990}]{Broadhurst90}
Broadhurst, T.J., Ellis, R.S., Koo, D.S., Szalay, A.S. 1990, 
Nature, 343, 726
\bibitem[\protect\citename{Broadhurst}{1999}]{Broadhurst99}
Broadhurst, T.J., Jaffe, A. H., 1999, ApJ, submitted
(astro-ph: 9904348)
\bibitem[\protect\citename{Bunn}{1997}]{Bunn97}
Bunn, E.F., White, M. 1997, ApJ, 480, 6
\bibitem[\protect\citename{Carlberg}{1997}]{Carlberg97}
Carlberg, R.G., Yee, H.K.C., Ellingson, E. 1997, ApJ, 479, L19
\bibitem[\protect\citename{Caztanaga}{1998}]{Caztanaga98}
Caztanaga, E., Baugh, C.M. 1998, MNRAS, 294, 229
\bibitem[\protect\citename{Efstathiou}{1988}]{Efstathiou88}
Efstathiou, G., Frenk, C.S., White, S.D.M., Davis, M. 1988, 
MNRAS, 235, 715
\bibitem[\protect\citename{Einasto}{1997}]{Einasto97}
Einasto, J., et al. 1997, Nature, 385, 139
\bibitem[\protect\citename{Einasto}{1999}]{Einasto99}
Einasto, J., et al. 1999, ApJ, 519, 441
\bibitem[\protect\citename{Eisenstein}{1998}]{Eisenstein98}
Eisenstein, D.J., Hu, W., Silk, J., Szalay, A.S. 1998, ApJ, 494, L1
\bibitem[\protect\citename{Eke}{1996}]{Eke96}
Eke, V. R., Cole, S., Frenk, C. S. 1996, MNRAS, 282, 263
\bibitem[\protect\citename{Giovanelli}{1997}]{Giovanelli97}
Giovanelli, R. et al. 1997, AJ, 113, 22
\bibitem[\protect\citename{Girardi}{1998}]{Girardi98}
Girardi, M., Borgani, S., Giuricin, G., Mardirossian, F., Mezetti, M. 
1998, ApJ, 506, 45 (G98)
\bibitem[\protect\citename{Gramann}{1999}]{Gramann99}
Gramann, M., Suhhonenko, I. 1999, ApJ, 519, 433
\bibitem[\protect\citename{Kaiser}{1987}]{Kaiser87}
Kaiser, N., 1987, MNRAS, 227, 1
\bibitem[\protect\citename{Katgert}{1998}]{Katgert98}
Katgert, P., Mazure, A., den Hartog, R., Adami, C., Biviano, A., Perea,
J., 1998, A \& AS, 129, 399
\bibitem[\protect\citename{Kitayama}{1996}]{Kitayama96}
Kitayama, T., Suto, Y. 1996, ApJ, 469, 480
\bibitem[\protect\citename{Kolatt}{1997}]{Kolatt97}
Kolatt, T., Dekel, A., 1997, ApJ, 479, 592
\bibitem[\protect\citename{Lacey}{1994}]{Lacey94}
Lacey, C., Cole, S. 1994, MNRAS, 271, 676
\bibitem[\protect\citename{Landy}{1994}]{Landy96}
Landy, S.D., Shectman, S.A., Lin, H., Kirshner, R.P., Oemler, A.,
Tucker, D., Schechter, P.L. 1996, ApJ, 456, L1
\bibitem[\protect\citename{Lesqourgues}{1997}]{Lesgourgues97}
Lesgourgues, J., Polarski, D., Starobinsky, A.A., 1998, MNRAS, 
297, 769 (LPS)
\bibitem[\protect\citename{Navarro}{1996}]{Navarro96}
Navarro, J.F., Frenk, C.S, White, S.D.M. 1996, ApJ, 462, 563
\bibitem[\protect\citename{Press}{1974}]{Press74}
Press, W.H., Schechter, P. 1974, ApJ, 187, 425
\bibitem[\protect\citename{Retzlaff}{1998}]{Retzlaff98}
Retzlaff, J., Borgani, S., Gottl\"ober, S., Klypin, A., M\"uller, V. 1998,
New A., 3, 631
\bibitem[\protect\citename{Seljak}{1996}]{Seljak96}
Seljak, U., Zaldarriaga, M., 1996, ApJ, 469, 7
\bibitem[\protect\citename{Starobinsky}{1992}]{Starobinsky92}
Starobinsky, A.A., 1992, JETP Lett., 55, 477
\bibitem[\protect\citename{Suhhonenko}{1999}]{Suhhonenko99}
Suhhonenko, I., Gramann, M., 1999, MNRAS, 303, 77
\bibitem[\protect\citename{Tammann}{1998}]{Tammann98}
Tammann, G., 1998, in General Relativity, 8th Marcel Crossmann
Symposium, ed. T. Piran, (Singapore:World Scientific)
\bibitem[\protect\citename{Watkins}{1997}]{Watkins97}
Watkins, R. 1997, MNRAS, 292, L59
\bibitem[\protect\citename{White}{1993}]{White93}
White, S. D. M., Efstathiou, G., \& Frenk, C. S. 1993, MNRAS, 
262, 1023
\bibitem[\protect\citename{Willick}{1997}]{Willick97}
Willick, J.A., Courteau S., Faber, S. M., Burstein, D., Dekel, A.,
Strauss, M. A., 1997, ApJS, 109, 333

\end{thebibliography}
\end{document}